\RequirePackage{lineno}
\documentclass[twocolumn,showpacs,aps,prd]{revtex4-1}

\usepackage{amssymb}
\usepackage{graphicx}
\usepackage{dcolumn}
\usepackage{bm}
\usepackage{epsfig}
\usepackage{amsfonts}
\usepackage{keyval,graphicx}
\usepackage{textcomp,wasysym}
\usepackage{subfigure}
\usepackage{hyphenat}
\usepackage{xcolor}

\long\def\inst#1{\par\nobreak\kern 4pt\nobreak
    {\itshape #1}\par\vskip 10pt plus 3pt minus 3pt}
\RequirePackage{xspace}
\usepackage{relsize}

\begin{document}

\title{\Large \boldmath \bf  Search for $Z_c(3900)^{\pm}\to\omega\pi^{\pm}$}
\author{
  \begin{small}
    \begin{center}
      M.~Ablikim$^{1}$, M.~N.~Achasov$^{9,f}$, X.~C.~Ai$^{1}$,
      O.~Albayrak$^{5}$, M.~Albrecht$^{4}$, D.~J.~Ambrose$^{44}$,
      A.~Amoroso$^{48A,48C}$, F.~F.~An$^{1}$, Q.~An$^{45,a}$,
      J.~Z.~Bai$^{1}$
% ...
      \\
      \vspace{0.2cm}
      (BESIII Collaboration)\\
      \vspace{0.2cm} {\it
        $^{1}$ Institute of High Energy Physics, Beijing 100049, People's Republic of China\\
% ...
$^{4}$ Bochum Ruhr-University, D-44780 Bochum, Germany\\
        $^{5}$ Carnegie Mellon University, Pittsburgh, Pennsylvania 15213, USA\\
% ...
$^{9}$ G.I. Budker Institute of Nuclear Physics SB RAS (BINP), Novosibirsk 630090, Russia\\
% ...
$^{44}$ University of Rochester, Rochester, New York 14627, USA\\
        $^{45}$ University of Science and Technology of China, Hefei 230026, People's Republic of China\\
% ...
$^{48}$ (A)University of Turin, I-10125, Turin, Italy; (B)University of Eastern Piedmont, I-15121, Alessandria, Italy; (C)INFN, I-10125, Turin, Italy\\
% ...
$^{a}$ Also at State Key Laboratory of Particle Detection and Electronics, Beijing 100049, Hefei 230026, People's Republic of China\\
% ...
$^{f}$ Also at the Novosibirsk State University, Novosibirsk, 630090, Russia\\
% ...
      }
    \end{center}
    \vspace{0.4cm}
  \end{small}
}

\affiliation{}


\begin{abstract}

The decay $Z_c(3900)^\pm\to\omega\pi^\pm$ is searched for using data
samples collected with the BESIII detector operating at the BEPCII
storage ring at center-of-mass energies $\sqrt{s}=4.23$ and
$4.26$~GeV.  No significant signal for the $Z_c(3900)^\pm$ is found, and
upper limits at the 90\% confidence level on the Born cross
section for the process $e^+e^-\to
Z_c(3900)^\pm\pi^\mp\to\omega\pi^+\pi^-$ are determined to be $0.26$ and
$0.18$ pb at $\sqrt{s}=4.23$ and 4.26 GeV, respectively.

\end{abstract}


\maketitle

\section{Introduction}

Recently, in the study of $e^+e^-\to J/\psi \pi^+\pi^-$, a distinct
charged structure, named the $Z_c(3900)^{\pm}$, was observed in the
$J/\psi\pi^\pm$ spectrum by BESIII~\cite{Ablikim:2013mio} and
Belle~\cite{Liu:2013dau}.  Its existence was confirmed shortly thereafter with CLEO-c
data~\cite{Xiao:2013iha}. The existence of the neutral partner in the
decay $Z_c(3900)^0\to J/\psi\pi^0$ has also been reported in CLEO-c
data~\cite{Xiao:2013iha} and by BESIII~\cite{BESIII:2015kha}. The
$Z_c(3900)$ is a good candidate for an exotic state beyond simple quark models,
since it contains
a $c\bar{c}$ pair and is also electrically charged. Noting that the
$Z_c(3900)$ has a mass very close to the $D^*\bar{D}$ threshold
(3875~MeV), BESIII analyzed the process
$e^{+}e^{-}\to\pi^{\pm}(D\bar{D}^{*})^{\mp}$, and a clear structure in
the $(D\bar{D}^{*})^{\mp}$ mass spectrum is seen, called the $Z_c(3885)$. The
measured mass and width are $(3883.9 \pm 1.5 \pm 4.2)$~MeV/$c^2$ and
$(24.8\pm 3.3 \pm 11.0)$~MeV, respectively, and quantum numbers $J^P =
1^+$ are favored~\cite{Ablikim:2013xfr}. Assuming the $Z_c(3885)\to
D\bar{D}^{*}$ and the $Z_c(3900)\to J/\psi\pi$ signals are from the
same source, the ratio of partial widths $\frac{\Gamma(Z_c(3885)\to
  D\bar{D}^{*})}{\Gamma(Z_c(3900)\to J/\psi\pi)}$ is determined to be $6.2\pm1.1\pm2.7$.

The observation of the $Z_c(3900)$ has stimulated many theoretical
studies of its nature. Possible interpretations are tetra-quark~\cite{Faccini:2013lda}, hadro-charmonium
\cite{Voloshin:2013dpa}, $D^*\bar{D}$ molecule~\cite{Guo:2013sya} and
threshold effects~\cite{ZhaoQ, LiuX, Swanson:2014tra}.
Lattice QCD studies provide theoretical support for the existence of
$X(3872)$ \cite{Prelovsek:2013cra} but not for the
$Z_c(3900)$~\cite{Prelovsek:2013xba, Prelovsek:2014swa,
Leskovec:2014gxa, Lee:2014uta,Chen:2014afa}.  However, those studies
were carried out on small volumes with unphysically heavy up and down
quarks. It is also worth noting that no resonant structure in $J/\psi\pi$
is observed in $\overline{B}^0\to J/\psi \pi^+\pi^-$ by LHCb
\cite{Aaij:2014siy}, in $\overline{B}^0\to J/\psi K^-\pi^+$ by Belle
\cite{Chilikin:2014bkk} or in $\gamma p\to J/\psi\pi^+ n$ by COMPASS
\cite{Adolph:2014hba}.

The decay properties of a state can provide useful information on its
internal structure. There are three important decay modes for
charmonium-like states: (i) ``fall-apart'' decays to open charm
mesons; (ii) cascades to hidden charm mesons; and (iii) decays to
light hadrons via intermediate gluons. In addition, as shown in
Ref.~\cite{ZhaoQ, LiuX}, an enhancement near the $D\bar{D}^*$ threshold
can be produced via rescattering of hidden or open charm final
states. Decays of the $Z_c(3900)$ to light hadrons can play a unique role
in distinguishing a resonance from threshold effects, because the
decay mode with $c\bar{c}$ annihilation involves neither hidden nor
open charm final states.  However, theory estimates of annihilation
widths to light hadrons are only order of magnitude due to uncertainties of
wave function effects and QCD
corrections~\cite{Close:2003mb,Braaten:2003he}.
A sizeable $Z_c(3900)$ decay width to light hadrons might be expected
in analogy to $\eta_c$ or $\chi_{cJ}$ into hadronic final states.

Among a large number of hadronic final states that are available for a
$I^G(J^P)=1^+(1^{+})$ resonance decay, $\omega\pi$ is one of the
typical decay modes which are not suppressed by any known selection
rule. In this paper, we report a search for $Z_c(3900)^\pm\to \omega
\pi^\pm$ based on $e^+e^-$ annihilation samples taken at
center-of-mass (CM) energies $\sqrt{s}=4.23$ and 4.26 GeV. The data
samples were collected with the BESIII~\cite{bes3dect} detector
operating at the BEPCII storage ring. The integrated luminosity of
these data samples are measured by analyzing the large-angle Bhabha
scattering events with an uncertainty of 1.0\%~\cite{lum} and are equal to
1092~pb$^{-1}$ and 826~pb$^{-1}$, for $\sqrt{s}=4.23$ and 4.26~GeV,
respectively.


\section{BESIII EXPERIMENT AND MONTE CARLO SIMULATION}

The BESIII detector, described in detail in Ref.~\cite{bes3dect}, has a
geometrical acceptance of 93\% of 4${\pi}$. A small-cell helium-based
main drift chamber (MDC) provides a charged particle momentum
resolution of 0.5\% at 1 GeV/$c$ in a 1 T magnetic field, and supplies
energy-loss ($dE/dx$) measurements with a resolution of 6\%
for minimum-ionizing pions. The electromagnetic calorimeter
(EMC) measures photon energies with a resolution of 2.5\% (5\%) at 1.0
GeV in the barrel (end-caps). Particle identification (PID) is provided
by a time-of-flight system (TOF) with a time resolution of 80 ps (110
ps) for the barrel (end-caps). The muon system, located in the iron
flux return yoke of the magnet, provides 2 cm position resolution and
detects muon tracks with momenta greater than 0.5 GeV/$c$.

The {\sc geant}4-based~\cite{Agostinelli:2002hh} Monte Carlo (MC)
simulation software {\sc boost}~\cite{boost} includes the geometric
description of the BESIII detector and a simulation of the detector
response.  It is used to optimize event selection criteria, estimate
backgrounds and evaluate the detection efficiency.  We generate signal
MC samples of $e^+e^-\to Z_c(3900)^\pm\pi^\mp\to\omega\pi^+\pi^-$
uniformly in phase space, where the $\omega$ decays to
$\pi^+\pi^-\pi^0$.
The decays of $\omega\to\pi^+\pi^-\pi^0$ are generated with the
\textit{OMEGA$\_$DALITZ} model in \textsc{evtgen}~\cite{GEN, bes3gen}.
Initial state radiation (ISR) is simulated with
\textsc{kkmc}~\cite{kkmc2000,kkmc2001}, where the Born cross section
of $e^+e^-\to Z_c(3900)^\pm\pi^\mp$ is assumed to follow a $Y(4260)$
Breit-Wigner (BW) line shape with resonance parameters taken from the
Particle Data Group (PDG)~\cite{PDG}.  Final state radiation (FSR)
effects associated with charged particles are handled with
\textsc{PHOTOS}~\cite{kkmc2000}.  For studies of possible backgrounds,
inclusive $Y(4260)$ MC samples with luminosity equivalent to the
experimental data at $\sqrt{s}=4.23$ and $\sqrt{s}=4.26$~GeV are
generated, where the main known decay channels are generated using
\textsc{evtgen}~\cite{GEN, bes3gen} with branching fractions taken
from the PDG~\cite{PDG}. The remaining events associated with charmonium
decays are generated with \textsc{lundcharm}~\cite{lundcharm}, while
continuum hadronic events are generated with
\textsc{PYTHIA}~\cite{Sjostrand:2006za}. QED processes such as Bhabha
scattering, dimuon and digamma events are generated with \textsc{kkmc}~\cite{kkmc2000,kkmc2001}.

\section{Data analysis and Background study}

Tracks of charged particles in BESIII are reconstructed from MDC
hits. We select tracks with their point of closest approach within
$\pm 10$~cm of the interaction point in the beam direction and within
$1$~cm in the plane perpendicular to the beam. Information from the
TOF and $dE/dx$ measurements
are combined to form PID confidence levels for the $\pi$
and $K$ hypotheses; each track is assigned to the particle type with
the highest confidence level.

Photon candidates are reconstructed by clustering EMC crystal
energies. The efficiency and energy resolution are improved by
including energy deposits in nearby TOF counters. The minimum energy
is required to be $25$~MeV for barrel showers ($|\cos\theta| < 0.80$)
and $50$~MeV for endcap showers ($0.86<|\cos\theta|<0.92$). To exclude
showers from charged particles, the angle between the shower and the
extrapolated charged tracks at the EMC must be greater than
$5^{\circ}$. A requirement on the EMC cluster timing with respect to the event start time
is applied to suppress electronic noise and energy
deposits unrelated to the event.

The $\pi^{0}$ candidates are formed from pairs of photons that can
be kinematically fitted to the known $\pi^0$ mass. The $\chi^{2}$
from this fit with one degree of freedom is required to be less than
$25$.

Events with exactly four charged tracks identified as pions with zero net
charge and at least one $\pi^{0}$ candidate are selected.  A
five-constraint kinematic fit (5C) is performed to the hypothesis of
$e^+e^-\to \pi^{+}\pi^{-}\pi^{+}\pi^{-}\pi^{0}$ (constraints are the
4-momentum of the initial $e^+e^-$ system and the $\pi^{0}$ mass), and
$\chi^{2}_{5C}<40$ is required. If there more than one $\pi^{0}$ is
found in an event,
the combination with the smallest $\chi^{2}_{5C}$ is retained.

\begin{figure}[htbp]

\includegraphics[angle=0,width=8cm, height=6cm]{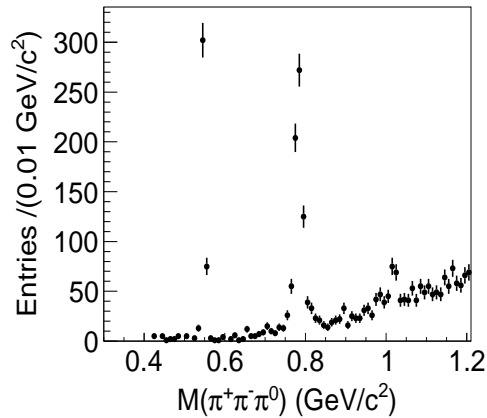}
\caption{The $\pi^+\pi^-\pi^{0}$ invariant mass distribution of the
  combination closest to the $\omega$, for the selected $e^+e^-\to
  \pi^{+}\pi^{-}\pi^{+}\pi^{-}\pi^{0}$ candidates for the data sample at
  $\sqrt{s}=4.23$~GeV.}
\label{fig_3pi}
\end{figure}

Figure~\ref{fig_3pi} shows the $\pi^+\pi^-\pi^{0}$ invariant mass
distribution of the $\pi^+\pi^-\pi^{0}$ combination with invariant
mass closest to the mass of $\omega$ for the selected candidate
$e^+e^-\to \pi^{+}\pi^{-}\pi^{+}\pi^{-}\pi^{0}$ events at
$\sqrt{s}=4.23$~GeV, where prominent $\eta$, $\omega$ and $\phi$
signals are observed.

\begin{figure}[htbp]

\includegraphics[angle=0,width=8.0cm, height=6.0cm]{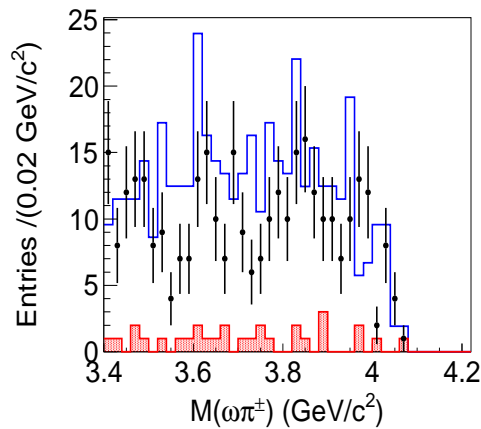}
\caption{ Distribution of $M(\omega\pi^{\pm})$ for the data sample
  at $\sqrt{s}=4.23$~GeV.  The dots with error bars are events
  within the $\omega$ signal region. The shaded histogram shows events
  selected from the $\omega$ sidebands, and the solid histogram shows
  inclusive MC events, which are dominated by continuum events.}
\label{fig_omegapi}
\end{figure}

$\omega$ candidates are selected with the mass window
$|M(\pi^+\pi^-\pi^{0})_\mathrm{closest}-m_\omega|<$0.03~GeV$/c^{2}$, where
$m_\omega$ is the nominal mass of the $\omega$ taken from the
PDG~\cite{PDG}. Figure~\ref{fig_omegapi} shows the $M(\omega \pi^\pm)$
distribution for the candidate events of $e^+e^-\to
\omega\pi^{+}\pi^{-}$ at $\sqrt{s}=4.23$~GeV. No sign of a peak near
$3.9$~GeV$/c^{2}$ is apparent. The shaded histogram in
Fig.~\ref{fig_omegapi} shows the distribution of non-$\omega$
background for the events in $\omega$ sideband regions
($0.06<|M(\pi^+\pi^-\pi^{0})_\mathrm{closest}-m_\omega|<0.09$~GeV$/c^{2}$).

By studying inclusive MC samples with luminosity equivalent to the
data at $\sqrt{s}=4.23$ and 4.26~GeV, the background is found to be
dominantly from the continuum process $e^+e^-\to
\omega\pi^{+}\pi^{-}$. The solid histogram in Fig.~\ref{fig_omegapi}
shows the $\omega\pi^\pm$ invariant mass distribution for events
selected from the inclusive MC sample.

\section{Fitting Results}\label{fit}

We use a one-dimensional, unbinned, extended maximum likelihood fit to
the $\omega\pi^\pm$ invariant mass distribution to obtain
the yield of $Z_c(3900)^\pm\to\omega\pi^\pm$ events. The signal
probability density function (PDF) is parameterized by an
$\mathcal{S}$-wave Breit-Wigner function convolved
with a Gaussian resolution function and weighted with the detection efficiency:
\begin{eqnarray}
 \left(G(M;\sigma) \otimes \frac{p\cdot q}{(M^{2}-M^{2}_{0})^{2}+M^{2}_{0}\Gamma^{2}}
\right)\times \varepsilon (M) \ ,
\end{eqnarray}
where $G(M;\sigma)$ is a Gaussian function representing the mass resolution. The
mass resolution of the $Z_c(3900)^\pm$ is $1.2\pm 0.1$~MeV/$c^2$ at
both $\sqrt{s}=4.23$ and 4.26~GeV, according to MC simulation. $p\cdot
q$ is the $S$-wave phase space factor, where $p$ is the
$Z_c(3900)^\pm$ momentum in the $e^+e^-$ CM frame and $q$ is the
$\omega$ momentum in the $Z_c(3900)^\pm$ CM frame.  $M$ is the
invariant mass of $\omega\pi^\pm$, and $M_{0}$ and $\Gamma$ are the mass
and width of the $Z_c(3900)^\pm$, which are fixed to the results in
Ref~\cite{Ablikim:2013mio}. $\varepsilon (M)$ is the efficiency curve
as a function of the $\omega\pi^\pm$ invariant mass, obtained from
signal MC simulation.

The background shape is described by an ARGUS function
$M\sqrt{1-(M/m_0)^2}\cdot\exp(c(1-(M/m_0)^2))$, where $c$ is left
free in the fit and $m_0$ is fixed to the threshold of
$\sqrt{s}-m_{\pi}$~\cite{Albrecht:1990am}.

\begin{figure}[htbp]
\includegraphics[angle=0,width=8cm, height=6cm]{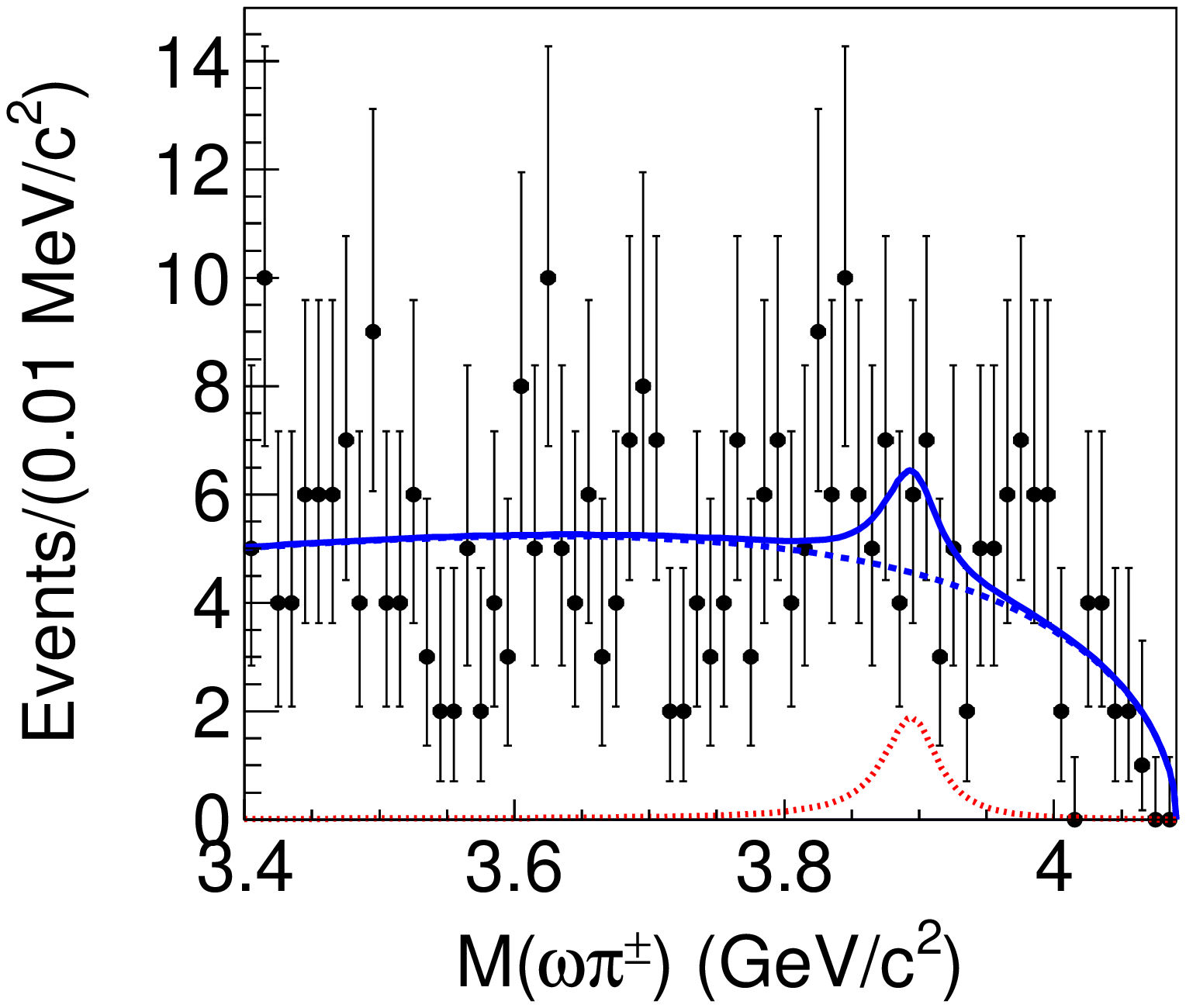} \put(-75,150){(a)}\\
\includegraphics[angle=0,width=8cm, height=6cm]{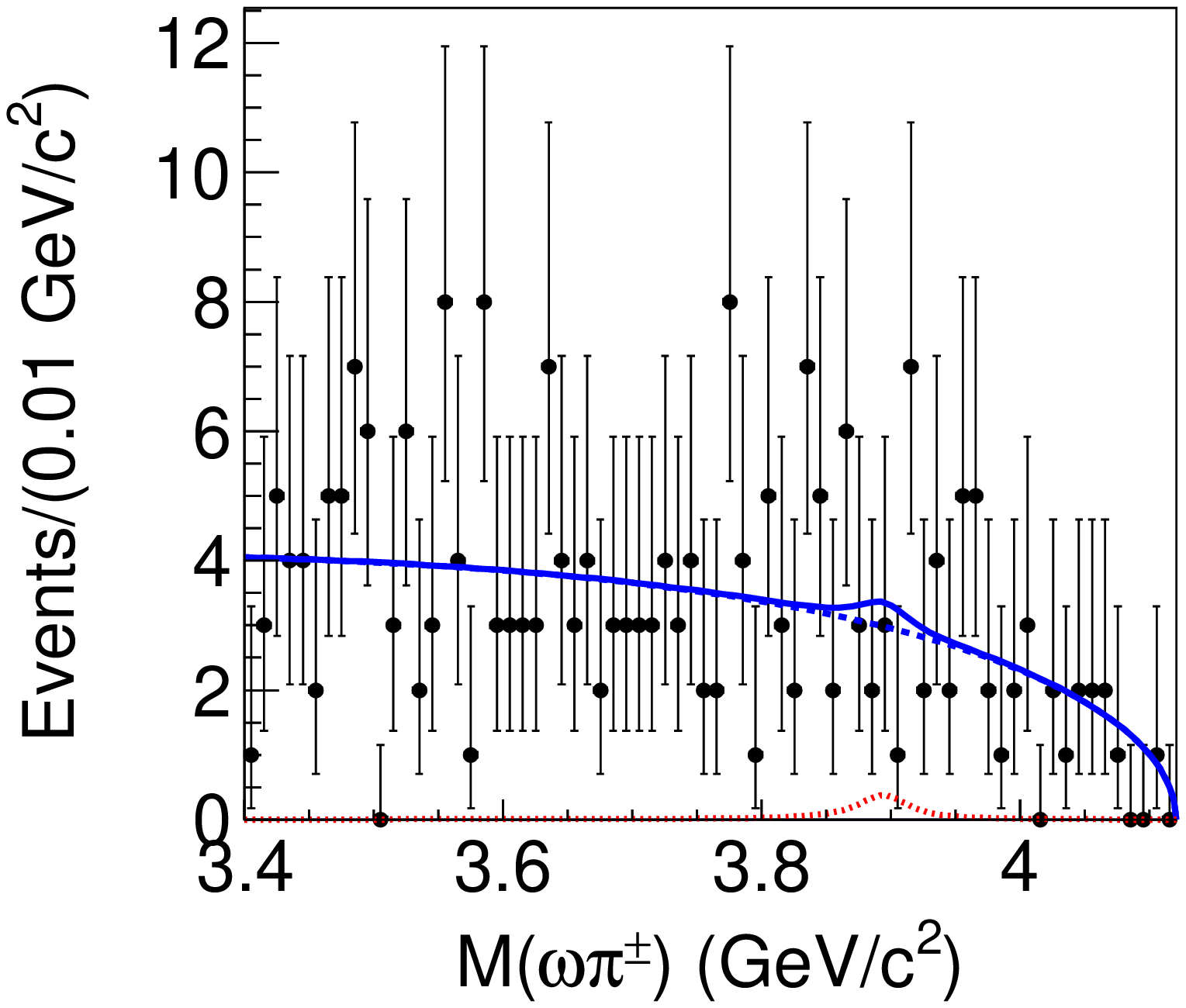} \put(-75,150){(b)}
\caption{Results of the unbinned maximum likelihood fit of the
  $\omega\pi^{\pm}$ mass spectrum of $e^+e^-\to\omega\pi^{+}\pi^{-}$
  at (a) $\sqrt{s}=4.23$~GeV and (b) $\sqrt{s}=4.26$~GeV.  Dots with
  error bars are the data. The solid curve is the result of the fit
  described in the text. The dotted curve is the $Z_c(3900)^\pm$
  signal. The dashed curve is the background.}
\label{fig_fit_4230}
\end{figure}

Figure~\ref{fig_fit_4230}(a) shows the fit result for the data
sample at $\sqrt{s}=4.23$~GeV.
The fit yields $14\pm11$ events for the $Z_c(3900)^\pm$
signal. Compared to the fit without the $Z_c(3900)^\pm$ signal, the
change in $\ln L$ with $\Delta(d.o.f.)=1$ is 0.74, corresponding to a
statistical significance of $1.2\sigma$. Using the Bayesian
method~\cite[Sect.38.4.1]{PDG}, the upper limit for the $Z_c(3900)^\pm$ signal is
set to $33.5$ events at the $90\%$ confidence level (C.L.), where only
the statistical uncertainty is considered.

The fit result for the data sample at $\sqrt{s}=4.26$~GeV is shown
in Fig.~\ref{fig_fit_4230}(b). The fit yields $2.2\pm8.1$ events for
the $Z_c(3900)^\pm$ with a statistical significance of
$0.1\sigma$. The upper limit is $18.8$ events at the $90\%$ C.L.

\section{Cross section upper limits and Systematic uncertainty}

The upper limit on the Born cross section at the $90\%$ C.L. is calculated as
\begin{eqnarray}\label{xsec}
   \sigma(e^+e^-\to Z_c(3900)^\pm\pi^\mp, Z_c(3900)^\pm\to\omega\pi^\pm)=\nonumber\\
   \frac{N^{\rm UL}}{\mathcal{L}_{\rm int}(1+\delta)\frac{1}{|1-\Pi|^{2}}\epsilon(1-\sigma_{\epsilon} )\mathcal{B}_\omega\mathcal{B}_{\pi^0}},
\end{eqnarray}
where $N^{\rm UL}$ is the upper limit on the signal events;
$\mathcal{L}_{\rm int}$ is the integrated luminosity; $\epsilon$ is
the selection efficiency obtained from signal MC simulation, which are
18.5$\pm$0.2\% and 18.6$\pm$0.2\% at $\sqrt{s}=4.23$ and $4.26$~GeV,
respectively; $\sigma_{\epsilon}$ is the systematic uncertainty of the
efficiency described in next paragraph; $\frac{1}{|1-\Pi|^{2}}$ is the
vacuum polarization factor obtained by using calculations from
Ref.~\cite{Actis:2010gg}, and equal to 1.06 for both energies;
$(1+\delta)$ is the radiative correction factor, equal to $0.844$ for
$\sqrt{s}=4.23$~GeV and 0.848 for $\sqrt{s}=4.26$~GeV obtained using
Ref.~\cite{kkmc2000,kkmc2001} by assuming the line shape of Born cross
section $\sigma(e^+e^-\to Z_c(3900)^\pm\pi^\mp)$ to be a BW function
with the parameters of the Y(4260) taken from PDG~\cite{PDG}; and
$\mathcal{B}_\omega$ and $\mathcal{B}_{\pi^0}$ are the branching fractions
of the decay $\omega\to\pi^+\pi^-\pi^0$ and
$\pi^0\to\gamma\gamma$~\cite{PDG}, respectively. A conservative
estimate of the upper limit of the Born cross section is determined by
lowering the efficiency by one standard deviation of the systematic
uncertainty.

The systematic uncertainty of the cross section measurement from
Eq.~\ref{xsec} is summarized in Table~\ref{table_syserr_1}.  The
luminosity is measured using Bhabha events with an uncertainty of
1.0\% ~\cite{lum}. The uncertainty in tracking efficiency for pions is
1.0\% per track~\cite{Ablikim:2013xfr}, i.e. 4.0\% for the track
selection in this analysis. The uncertainty in PID efficiency for
pions is 1.0\% per track~\cite{Ablikim:2013xfr}. The uncertainty in the
photon reconstruction efficiency is less than 1\% per
photon~\cite{Ablikim:2010zn}. The uncertainty in the $\pi^0$
reconstruction efficiency is 2.0\%~\cite{Ablikim:2010aa}.  The
uncertainty of the kinematic fit is estimated by correcting the helix
parameters of the charged tracks. The detailed procedure to extract the
correction factors can be found in Ref.~\cite{Ablikim:2012pg}. The
track parameters in MC samples are corrected by these factors, and the
difference in efficiencies of 0.8\% with and without the correction is
taken as the systematic uncertainty associated with the kinematic fit.  An MC
sample generated with $Z_c(3900)^\pm\to\omega\pi^\pm$ in both $S$ wave and
$D$ wave, assuming a $D/S$ waves amplitude ratio of 0.1, results in a
3\% change in detection efficiency. This difference is taken as the
systematic uncertainty associated with the MC production model.  The
branching ratio value for $\omega\to\pi^+\pi^-\pi^0$ comes from the
PDG~\cite{PDG}, and its error is 0.8\%. In the nominal fit, the radiative correction factor
and the detection efficiency are determined under the assumption that the
production of $e^+e^-\to Z_c(3900)^\pm\pi^\mp$ follows the $Y(4260)$ line
shape. Using the line shape of $\sigma(e^+e^-\to
Z_c(3900)^0\pi^0)$ measured in Ref.~\cite{BESIII:2015kha} as an
alternative assumption, $\epsilon(1+\delta)$ is increased by 6\% for
$\sqrt{s}=4.23$~GeV and 7\% for $\sqrt{s}=4.26$~GeV . The change in
$\epsilon(1+\delta)$ is taken as a systematic uncertainty. The
uncertainty of the vacuum polarization factor is taken from
Ref.~\cite{Actis:2010gg}, and is negligible compared with other
uncertainties. Assuming that all sources of systematic uncertainties
are independent, the total errors are given by the quadratic sums.

\begin{table}
\caption{Summary of the relative systematic uncertainties of the cross section
  measurement (in \%).}
\label{table_syserr_1}
\begin{tabular}{ccc}
  \hline\hline
  Source     & $\sqrt{s}=4.23$~GeV & $\sqrt{s}=4.26$~GeV           \\
  \hline
  Luminosity         & 1.0& 1.0  \\
  Tracking           & 4.0 & 4.0 \\
  PID           & 4.0& 4.0 \\
  photon reconstruction  & 2.0 & 2.0  \\
  $\pi^0$ reconstruction  & 2.0 & 2.0  \\
  Kinematic fit      & 0.8& 0.8  \\
  Decay model        & 3& 3 \\
  Radiative correction &6&7\\
  $Br(\omega\to\pi^+\pi^-\pi^0)$   & 0.8& 0.8 \\

 Total              & 9.4 & 10.1 \\
  \hline\hline
\end{tabular}
\end{table}

To estimate the systematic uncertainties due to the fit procedure,
we fit under different scenarios, and the upper limits obtained at the
90\% C.L. for the $Z_c(3900)^\pm$ signal yield are summarized in
Table~\ref{table_syserr_2}.  The effect on the signal yield from the
fit range is obtained by varying the fit range by
$\pm0.1$~GeV/$c^2$.  The effect due to the choice of the background
shape is estimated by changing the background shape from the ARGUS
function to a second order polynomial (where the parameters of the
polynomial are allowed to vary and the fit range is limited to
[3.4, 4.08]~GeV/$c^2$). The effect due to the resonance parameters of
the $Z_c(3900)^\pm$ is estimated by varying the resonance parameters
according to the results in Ref~\cite{Ablikim:2013xfr}.  The effect
due to the mass resolution is estimated by increasing the resolution
by 10\% according to the comparison between the data and MC. The
effect due to the mass-dependent efficiency curve is estimated by changing the
efficiency curve to a constant function. We take the largest number of
$Z_c(3900)^\pm$ events in the different scenarios as a conservative
estimate of the upper limit: $N^{\rm UL}_{4230}= 38.0$, $N^{\rm UL}_{4260}=
18.8$. The resulting upper limits of the Born cross sections at
$\sqrt{s}=4.23$ and $4.26$~GeV are determined to be 0.26 and 0.18~pb
at the $90\%$ C.L., respectively.

\begin{table}
\caption{Results of upper limits on the $Z_c(3900)$ signal yield with various fit procedures.}
\label{table_syserr_2}
\begin{tabular}{ccc}
  \hline\hline
  Source      & $\sqrt{s}=4.23$~GeV & $\sqrt{s}=4.26$~GeV         \\
  \hline
  Fit range         & 31.5 & 18.5 \\
  Background shape           & 38.0 & 16.1 \\
  $Z_c(3900)$ mass and width           & 22.6 & 12.2\\
  Mass resolution  & 33.5 & 18.8\\
  Efficiency curve      & 33.3 & 18.8\\
  \hline\hline
\end{tabular}
\end{table}

\section{Summary and Discussion}

In summary, based on data samples of 1092 pb$^{-1}$ at
$\sqrt{s}=4.23$~GeV and 826 pb$^{-1}$ at $\sqrt{s}=4.26$ GeV collected
with the BESIII detector operating at the BEPCII storage ring, a search is
performed for the decay $Z_c(3900)^\pm\to\omega\pi^\pm$ in $e^+e^-\to
\omega\pi^{+}\pi^{-}$. No $Z_c(3900)^\pm$ signal is observed. The
corresponding upper limits on the Born cross section are set to be
0.26 and 0.18 pb at $\sqrt{s}=4.23$ and 4.26 GeV, respectively.  If we
assume that the $Z_c(3900)^\pm$ observed in $e^+e^-\to
J/\psi\pi^{+}\pi^{-}$~\cite{Ablikim:2013mio} and $Z_c(3885)^\pm$ in
$e^+e^-\to (D\bar{D}^{*})^\pm\pi^\mp$~\cite{Ablikim:2013xfr} are the
same particle, the decay width of $Z_c(3900)^\pm\to \omega \pi^\pm$ is
estimated to be smaller than 0.2\% of the $Z_c(3900)^\pm$ total width.  As
$\omega \pi$ is a typical light hadron decay mode of a
$I^G(J^P)=1^+(1^{+})$ resonance, the non-observation of
$Z_c(3900)^\pm\to\omega\pi^\pm$ may indicate that the annihilation of
$c\bar{c}$ in $Z_c(3900)^\pm$ is suppressed. Complementary to the
searches for $Z_c(3900)$ production \cite{Aaij:2014siy,
  Chilikin:2014bkk, Adolph:2014hba}, exploring new $Z_c(3900)$ decay
modes may provide a significant input to clarify its dynamical origin.

\acknowledgements

The BESIII collaboration thanks the staff of BEPCII and the IHEP
computing center for their strong support. This work is supported in
part by National Key Basic Research Program of China under Contract
No. 2015CB856700; National Natural Science Foundation of China (NSFC)
under Contracts Nos. 11125525, 11235011, 11322544, 11335008, 11425524;
the Chinese Academy of Sciences (CAS) Large-Scale Scientific Facility
Program; Joint Large-Scale Scientific Facility Funds of the NSFC and
CAS under Contracts Nos. 11179007, U1232201, U1332201; CAS under
Contracts Nos. KJCX2-YW-N29, KJCX2-YW-N45; 100 Talents Program of CAS;
INPAC and Shanghai Key Laboratory for Particle Physics and Cosmology;
German Research Foundation DFG under Contract No. Collaborative
Research Center CRC-1044; Istituto Nazionale di Fisica Nucleare,
Italy; Ministry of Development of Turkey under Contract
No. DPT2006K-120470; Russian Foundation for Basic Research under
Contract No. 14-07-91152; U. S. Department of Energy under Contracts
Nos. DE-FG02-04ER41291, DE-FG02-05ER41374, DE-FG02-94ER40823,
DESC0010118; U.S. National Science Foundation; University of Groningen
(RuG) and the Helmholtzzentrum fuer Schwerionenforschung GmbH (GSI),
Darmstadt; WCU Program of National Research Foundation of Korea under
Contract No. R32-2008-000-10155-0.

\end{document}